\documentclass[12pt,preprint]{aastex}
\usepackage{psfig}

\begin{document}

%%%%%%%%%%%%%% MY DEFINITIONS
\def\pdot {\dot P}
\def\Omdot {\dot \Omega}
\def\ltsima{$\; \buildrel < \over \sim \;$}
\def\lsim{\lower.5ex\hbox{\ltsima}}
\def\gtsima{$\; \buildrel > \over \sim \;$}
\def\gsim{\lower.5ex\hbox{\gtsima}}
\def\msole{~M_{\odot}}
\def\mdot {\dot M}
\def\gr {GRB~030227~}
\def\cha {\textit{Chandra~}}
\def\xmm  {\textit{XMM-Newton~}}
%%%%%%%%%%%%%%%%%%%%%%%%%%%%%%%%%%%%%%

\title{INTEGRAL and XMM-Newton observations of the weak GRB 030227
\footnote{Based on observations with INTEGRAL, an ESA project with instruments and
science data centre funded by ESA member states (especially
the PI countries: Denmark, France, Germany, Italy, Switzerland, Spain),
Czech Republic and Poland, and with the participation of Russia
and the USA, and \xmm an ESA science mission with instruments
and contributions directly  funded by ESA member states and the USA.}
% \\
%Draft 1.0 19/3/03 Confidential
}
\author{S.Mereghetti, D.G\"{o}tz, A.Tiengo}
\affil{Istituto di Astrofisica Spaziale e Fisica Cosmica, \\
Sezione di Milano  ''G.Occhialini'' - CNR v.Bassini 15, I-20133 Milano, Italy, \\
sandro, diego, tiengo @mi.iasf.cnr.it}
\author{V. Beckmann\altaffilmark{1}
\altaffiltext{1}{Institut f\"ur Astronomie und Astrophysik, Universit\"at T\"ubingen, Sand 1, D-72076 T\"ubingen, Germany},
J. Borkowski, T.J.-L. Courvoisier}
\affil{INTEGRAL Science Data Center, Chemin d' \'Ecogia 16, CH-1290 Versoix, Switzerland}
\author{A. von Kienlin, V. Schoenfelder}
\affil{Max-Planck-Institut f\"{u}r extraterrestrische Physik, \\
Postfach 1312 D-85741, Garching, Germany}
\author{J.P. Roques, L. Bouchet}
\affil{CESR, Toulouse, France }
\author{P. Ubertini}
\affil{Istituto di Astrofisica Spaziale e Fisica Cosmica, Roma - CNR }
\author{A. Castro-Tirado}
\affil{Instituto de Astrof\'{\i}sica de Andaluc\'{\i}a (IAA-CSIC), P.O. Box 03004,
18080 Granada, Spain\\}
\author{F. Lebrun, J. Paul}
\affil{DAPNIA/Service d'Astrophysique, Saclay, France}
\author{N. Lund}
\affil{Danish Space Research Institute, Copenhagen, Denmark }
\author{M. Mas Hesse}
\affil{ Centro de Astrobiologa (CSIC-INTA) }
\author{W. Hermsen, P. den Hartog}
\affil{SRON-National Institute for Space Research, Utrecht, The Netherlands }
\author{C. Winkler}
\affil{ESTEC, RSSD, Keplerlaan 1, 2201 AZ Nordwijk, The Netherlands  }

\begin{abstract}

We present INTEGRAL and \xmm observations of the prompt $\gamma$-ray emission and the
X--ray afterglow of \gr , the first GRB for which the quick localization
obtained with the INTEGRAL Burst Alert System (IBAS)  has
led to the discovery of X--ray and optical afterglows.
\gr had a duration of about 20 s and a peak flux of
$\sim$1.1 photons cm$^{-2}$ s$^{-1}$ in the 20-200 keV energy range.
The time averaged spectrum can be fit by a single  power law with
photon index  $\sim$2 and we find some evidence for a  hard to soft
spectral evolution.
The X--ray afterglow has been detected
starting only 8 hours after the prompt emission, with a
0.2-10 keV  flux decreasing as t$^{-1}$ from 1.3$\times$10$^{-12}$
to 5$\times$10$^{-13}$ erg cm$^{-2}$ s$^{-1}$.
The afterglow spectrum is well described by a power law
with photon index 1.94$\pm$0.05 modified
by a redshifted neutral absorber with column density of several  10$^{22}$ cm$^{-2}$.
A possible  emission line at  1.67 keV could be due to Fe
for a redshift $z\sim$3, consistent with the value inferred from the absorption.

\end{abstract}


\section{Introduction}

Our understanding of Gamma Ray Bursts (GRBs),
one of the mysteries of high-energy astrophysics for
more than 25 years, advanced dramatically after the discovery of their X--ray,
optical and radio afterglows
(see, e.g., van Paradijs, Kouveliotou \& Wijers  2000).
Currently, the multi-wavelengths study of GRBs is providing a wealth of results
relevant for several branches of astrophysics.
Besides the study of the prompt emission,
a rapid and accurate localization is crucial to pursue these
objectives and fully exploit the information on the GRB afterglows
and host environment.
Such a capability was first achieved by the \textit{BeppoSAX} satellite,
which located more than 45 GRBs from 1997 to 2002 (see, e.g., Amati et al. 2002),
and is currently provided by HETE II,
a satellite specifically devoted to this task (Ricker et al. 2002).

The INTEGRAL satellite
(\textit{INTErnational Gamma Ray Astrophysical Laboratory}, Winkler et al. 1999)
was successfully launched on October 17, 2002.
The INTEGRAL payload consists  of two
$\gamma$-ray instruments operating in the $\sim$15 keV - 10 MeV range:
% \footnote{INTEGRAL also carries two monitoring instruments for observations in the
% X--ray and optical bands (Lund et al. 1999, Gim\'enez et al. 1999),
% but \gr was located outside their fields of view.}
%on board INTEGRAL:
IBIS (Ubertini et al. 1999) and SPI (Vedrenne et al. 1999).
Both are coded mask telescopes\footnote{With the coded mask technique
it is possible to obtain images at energies where photons cannot be easily
reflected by interposing a partially absorbing mask between the source and a
position sensitive detector (see, e.g., Dean 1983).}
optimized for   angular and spectral resolution respectively.
%In addition INTEGRAL carries two monitoring instruments, JEM-X and OMC,
%for observations in the X--ray and optical bands respectively
%(Lund et al. 1999, Gim\'enez et al. 1999).

Although not specifically devoted to GRB studies, thanks to the large
field of view and good imaging capabilities of  its $\gamma$-ray instruments,
INTEGRAL is able to localize GRBs at an expected rate of 1-2 per month
(Mereghetti, Cremonesi \& Borkowski 2001).
%During the observations
The data are immediately transmitted to the ground and it is
possible to derive the coordinates of   detected GRBs
with very small delays. Automatic software, the INTEGRAL Burst Alert
System (IBAS, Mereghetti et al. 2001), has been developed at the
INTEGRAL Science Data Center (ISDC, Courvoisier et al. 1999) to
exploit these capabilities.
As soon as   INTEGRAL telemetry reaches the ISDC, the IBAS software screens
the data in  real time
looking for the presence of potential GRBs,   performs a rapid
imaging analysis of the candidates, and eventually distributes
the positions of the GRBs via the Internet.

In line with   pre-launch expectations,
five GRBs have been detected to date in the field of view of the INTEGRAL  instruments:
GRB021125 (Bazzano \& Paizis 2002),
GRB021219 (Mereghetti, G\"{o}tz \& Borkowski  2002),
GRB030131 (Borkowski et al. 2003), GRB030227 (G\"{o}tz et al. 2003a),
and GRB030320 (Mereghetti, G\"{o}tz \& Borkowski  2003).
Their positions have been  determined with errors in the 2$'$ to  4$'$ range.
Here we report on the observations of GRB030227,
the first   GRB for which a prompt localization with INTEGRAL has led to the
successful identification of X--ray  (Loiseau et al. 2003) and optical afterglows
(Castro-Tirado et al. 2003a, Soderberg et al. 2003).

\section{INTEGRAL observation}

IBIS  detected \gr with its low energy detector ISGRI
(Lebrun et al. 2001),
an array of 128$\times$128 CdTe crystals sensitive in the energy
range from $\sim$15 keV to  $\sim$300 keV.
ISGRI has an effective area of the order of 1000 cm$^{2}$ and provides
an angular resolution of $\sim$12$'$ over a 29$^{\circ}\times$29$^{\circ}$ field of view.
Bright sources can be located with   good  accuracy
(for example, the 90\% confidence level error radius for a source with
signal to noise ratio of $\sim$10 is as small as 1$'$).

The SPI instrument observes with  a coarser
angular resolution ($\sim$2$^{\circ}$) the same region of sky covered by IBIS,
providing  a better energy resolution (FWHM=2.5 keV at 1 MeV).
The detector consists of 19 Ge crystals cooled to
85 K, surrounded by a thick anticoincidence shield of BGO scintillation crystals.
%SPI is able to process photons which deposit their energy in a single detector
%(single events) or in several detectors (multiple events).
%The single events are furthermore analyzed on board by a pulse shape discriminator (PSD)
%which suppresses the background events due to $\beta$ decay.

\gr was detected by the IBAS program based on the analysis of the IBIS/ISGRI data,
consisting of positional, timing and energy information of each detected photon.
The IBAS alert message with the preliminary position of the burst
(R.A. = 4$^h$ 57$^m$ 24$^s$, Dec.= +20$^{\circ}$ 28$'$ 24$''$,  J2000)
was delivered to the members of the IBAS Team at  08:42:38 UT of February 27, 2003,
only 35 s after the start of the burst
(most of this delay was due to  buffering of the telemetry on board the satellite and to
data transmission between the ground station and the ISDC).
Unfortunately, the Internet message with these coordinates could not be
distributed in real time\footnote{The detection of \gr occurred during a calibration
observation of the Crab Nebula.
Since the  instrument configuration during these observations
caused some spurious IBAS triggers, the automatic delivery
of the alerts to external clients had been temporarily  disabled.
}.
Nevertheless, this information was distributed within less than one hour,
after the GRB had been confirmed by  an interactive quick look analysis
of the  data (G\"{o}tz et al. 2003a).
Further analysis resulted in an improved localization at
R.A. = 4$^h$ 57$^m$ 32.2$^s$,
Dec. = +20$^{\circ}$ 29$'$  54$''$
(G\"{o}tz et al. 2003b),
with  an error of 3$'$ dominated by systematic  uncertainties.
%which might still be present in the IBAS imaging procedures.
This position differs by  only 50$''$ from that of the optical transient
(Castro-Tirado et al. 2003).

The IBIS light curve of \gr   is shown in Fig. 1.
The burst, which  started approximately at 08:42:03 UT,
had a typical FRED
(Fast Rise and Exponential Decay) profile, with a rise to the peak in 2 s
and a decay well described by an exponential with  e-folding decay time  of 11$\pm$1 s.
The peak flux, integrated over 1 s,
is 1.1 photons cm$^{-2}$ s$^{-1}$ in the 20-200 keV energy range.
The fluence, in the same range and assuming the average spectrum discussed below, is
7.5$\times$10$^{-7}$ erg cm$^{-2}$.

\gr was detected  with a signal to noise ratio of $\sim$20
at off-axis angles Z=5.3$^{\circ}$, Y=6.9$^{\circ}$.
%  in the so called
%partially coded part of the field of view of IBIS. Only
%67\% of the detector area recorded the source flux modulated by the coded mask aperture.
Since a fully calibrated response
matrix valid at this off-axis angles is not yet available, we derived
the spectrum of \gr  by comparing its count rate
in different energy bins to the corresponding values obtained from the Crab Nebula
observed at a similar position in the field of view.
With this procedure we extracted the average burst spectrum
for the time interval 08:42:04  to 08:42:26 UT, which is
well described by a
single power law with photon index 1.85$\pm$0.2
over the 20-200 keV energy range.

\gr was also detected by SPI
%within the fully-coded field of view of SPI, which is slightly larger
%(diameter 16$^{\circ}$) than that of IBIS.
%It was detected
with a significance of  7.7 $\sigma$ in the 20-200 keV   range and localized
%It was therefore possible to obtain a
%good position
(R.A.= 74$^{\circ}$.766, Dec.= +20$^{\circ}$.531)
only 0.36$^{\circ}$   off  the IBIS position, owing to the less-accurate location
precision of SPI.
%The analysis of SPI data was performed by using
%single, PSD and multiple  events.
A spectrum was extracted for a time interval of 18 s starting at 08:42:04 UT.
The background was estimated  from a time interval of 35 min around
the GRB, but excluding the time span of the event itself.
A power law with  photon index of 1.9$\pm$0.3
and 20-200 keV flux of 4.7$\times$10$^{-8}$ erg cm$^{-2}$ s$^{-1}$
gives a good fit to these data. These values are consistent with those
obtained with IBIS, confirming  that the method used in the
IBIS spectral analysis does not introduce important systematic effects.

To study the spectral evolution as a function of time, we defined a hardness ratio
HR = (H-S)/(H+S)
based on the background subtracted  count rates
in the  ranges H=40-100 keV and S=20-40 keV.
From the HR values of both SPI and IBIS we found evidence for
a  hard to soft evolution.
%We therefore performed a
A time resolved spectral analysis gave the power law indices shown
in Fig. 2, which confirm the indication of a possible
spectral softening during the decaying part of the event.

%\gr was located outside the field of view of the OMC and JEM-X instruments.
The overall veto
count rate of SPI's anticoincidence shield (ACS) showed no evident count
rate increase which could be associated with the GRB event.
This is consistent with the fluxes quoted above and the
low effective area of the ACS for directions corresponding to  the SPI
field of view (von Kienlin et al. 2001).

\section{XMM-Newton observation}

\xmm observed the position of \gr\ for $\sim$13 hours,
starting on February 27 at 16:58 UT.
%for an observation length of  $\sim$13 hours.
Due to the presence of a high background induced by
low energy protons of solar origin, the observation had to be interrupted twice,
resulting in
net exposure times of 33 and 36 ks respectively in the PN and MOS cameras
of the EPIC instrument (Str\"{u}der et al. 2001, Turner et al. 2001).
All the cameras operated
in Full Frame mode and with the thin optical blocking filter.
The data were processed using SAS version 5.4.1.

A bright and variable source  was clearly detected at
R.A. = 04$^h$57$^m$ 33.1$^s$,
Dec. =  +20$^{\circ}$ 29$'$  05$''$ (error radius of 4$''$),
well  within the IBIS error circle of \gr.
The flux variability immediately suggested that
this source was the GRB afterglow
(Loiseau et al. 2003), as it was
later confirmed by the
discovery of a fading optical transient within its small error region.
Fig. 3 shows the background subtracted PN light curve
in the 0.2-10  keV energy range.
The figure also shows the first flux measurement (at $\sim$17:00 UT)
obtained with the  MOS camera, which started the observation earlier than the PN.
The X-ray flux decay  is well described by a power law
function,   F $\propto$ t$^{-\delta}$, with $\delta$=0.97$\pm$0.07.

The source spectra were extracted from a circle of radius 40$''$
and rebinned  to have at least 20 counts per channel and to
oversample by a factor three the instrumental energy resolution.
For the PN the background was extracted from a nearby rectangular region (1.5$'$$\times$2$'$)
in the same chip as the source and for the MOS cameras from an annulus
(inner and outer radii
of 1$'$ and 2$'$) centered on the source.
All the errors on the spectral parameters given below are at the 90\% confidence level.

After checking that no significant   variations in the best fit parameters,
except for the flux value, occurred during
the observation, we performed a spectral fit of the whole PN data set.
A model consisting of an absorbed power law resulted in an unacceptable
fit ($\chi^{2}$/dof=272/209) with
%of the longest time interval
%(from Feb 27 21:32 UT to Feb 28 06:07 UT) gave a
photon index $\Gamma$ = 2.04$\pm$0.05
and N$_{H}$ = (3.6$^{+0.3}_{-0.2})\times 10^{21}$ cm$^{-2}$,
%($\chi^{2}$/dof=272/209,
%This value of  N$_{H}$ is
larger than the galactic absorption in this direction
(N$_{H}$=1.75$\times$10$^{21}$ cm$^{-2}$, Hartmann \& Burton 1997).
An acceptable fit ($\chi^{2}$/dof=235/208, Fig. 4) could be obtained by fixing N$_{H}$
to the galactic value and adding a redshifted neutral absorption, N$_{Hz}$.
This resulted in
$\Gamma$ = 1.94$\pm$0.05,
N$_{Hz}$ =(6.8$^{+1.8}_{-3.8})\times10^{22}$ cm$^{-2}$,   $z$=3.9$\pm$0.3,
and 0.2-10 keV observed flux
F$_{x}$=8.5$\times10^{-13}$ erg cm$^{-2}$ s$^{-1}$.
%
%($\chi^{2}$/dof=286/270).
%The observed flux in the  0.2-10 keV range varied by a factor 2 during the
%PN observation (see Fig. 3).
%The average values was  F$_{x}$=???$\times10^{-13}$ erg cm$^{-2}$ s$^{-1}$.
As shown in Fig. 5, the redshift is not strongly constrained,
%by the fit results,
its value being correlated with that of   N$_{Hz}$.
%To investigate the dependence of t
The best fit values are only slightly dependent on the assumed galactic
%results on our assumption of the
%galactic absorption value, we repeated the fits for values of
N$_{H}$ value: by varying it by $\pm$30\% acceptable fits were always obtained with
$\Gamma\sim$2, while the 90\% c.l. range of $z$ and N$_{Hz}$ varied
by less than 35\%.

The fit residuals of Fig. 4 suggest the presence of possible lines in the
spectrum. We therefore tried to improve the fit by adding to the above model
gaussian lines at
different energies and fixed widths, smaller than the instrumental resolution.
The only possibly significant line was  found at  observed
energy of 1.67$^{+0.01}_{-0.03}$ keV.
%and 4.37$^{+0.06}_{-0.05}$ keV, with equivalent widths of
%37$^{+20}_{-15}$ eV and 180$^{+62}_{-76}$ eV
According to an F-test this line is significant at the 3.2 $\sigma$ level.
However, such a probability value should be used with caution (Protassov et al. 2002),
and we give it just to allow a comparison with other possible detections
of lines reported for previous GRB's.

%$\Gamma$ = 1.95$\pm 0.07$,  N$_{Hz}$ =
%(7.0$^{+2.2}_{-2.6})\times 10^{22}$ cm$^{-2}$, redshift  $z$=3.9$^{+0.2}_{-1.1}$,
%and
% observed flux F$_{x}$=7.9$\times10^{-13}$ erg cm$^{-2}$ s$^{-1}$
% in the 0.2-10 keV range.

% The same procedure applied to the PN data of the preceding time interval
% (from 18:27 to 20:32 UT), yielded entirely consistent values of the best fit parameters,
% except for a higher flux value, F$_{x}$=1.16$\times10^{-12}$ erg cm$^{-2}$ s$^{-1}$.
% We therefore proceeded to a joint fit of the two PN spectra (see Fig. 4), obtaining
% $\Gamma$ = 1.94$^{+0.06}_{-0.03}$,  N$_{Hz}$ =
% (6.7$^{+0.9}_{-3.7})\times10^{22}$ cm$^{-2}$ and   $z$=3.9$^{+0.3}_{-0.2}$
% ($\chi^{2}$/dof=286/270).
% As shown in Fig. 5, the redshift is not strongly constrained
% by the fit results, its value being correlated with that of   N$_{Hz}$.

We also tried thermal models,
both for the whole observation and for shorter time intervals,
replacing the power law component with
a redshifted optically thin  plasma model (MEKAL in XSPEC,
with $z$  linked  to that of N$_{Hz}$).
Keeping the elemental abundances fixed to the solar values resulted in
unacceptable fits.
% ($\chi^{2}$/dof=268/208).
Letting the abundances be free parameters,
we obtained slightly better results ($\chi^{2}$/dof=260/207),  with
values of N$_{Hz}$ and $z$ similar to the power law case, and a temperature
of $\sim$25 keV.
However, the corresponding  $\chi^{2}$/dof values were always greater than in
the power law model.
%, both for the whole observation ($\chi^{2}$/dof=260/208)
%and for the first and second time intervals separately.
%and for their sum).

Entirely consistent results
%for what concerns the continuum
were obtained from the MOS spectra.
%have enough statistics to
As mentioned above, a short initial exposure
was carried out only with the MOS, from 16:58 to 17:16 UT.
These data do not provide enough statistics for a detailed
spectral analysis, but they are
consistent with the spectral parameters obtained
for the rest of the observation.
The corresponding flux is 1.34$\times10^{-12}$ erg cm$^{-2}$ s$^{-1}$.

\section{Discussion}

The INTEGRAL data indicate that \gr belongs to the   class of long GRBs.
Its peak flux and fluence,
converted to the 50-300 keV energy range used in the BATSE catalog,
are respectively $\sim$0.6 photons cm$^{-2}$ s$^{-1}$ and
$\sim$7$\times$10$^{-7}$ erg cm$^{-2}$
(these values slightly depend on the extrapolation; for example,
a break in the spectrum at 120 keV to a slope of 2.5 would reduce them by 15\%).
About three quarter of the GRBs in the 4$^{th}$ BATSE Catalog (Paciesas et al. 1999)
have a higher peak flux, indicating that \gr is quite faint.
The possible evidence for a  hard to soft spectral evolution
in \gr suggests that  such a trend, already observed in samples of
brighter GRBs (Ford et al. 1995, Preece et al. 1998, Frontera et al. 2000), may
also apply to relatively faint bursts.

GRB spectra usually show a curvature requiring more complex models than a single
power law. A broken power law, with a smooth transition between the low
energy ($\alpha$) and high energy ($\beta$) slope provides in most cases
an adequate empirical description of the data (Band et al. 1993).
By fitting such a function to our   data no improvement of the fit was
obtained.
A spectral break, if any,   could occur outside the energy range
over which we detected \gr , most likely above $\sim$100-200 keV as seen in most
GRBs.
Our photon index $\sim$1.9 lies in the lower tail of the distribution of
values of $\alpha$ (Preece et al. 2000), indicating that \gr is relatively soft.
An extrapolation of its spectrum to lower energies, leads to a
ratio of fluences in the 6-30 to 30-400 keV ranges of
0.45 (or more if there is a high-energy break).
This would qualify \gr as an X--ray rich GRB
(see, e.g., Barraud et al. 2003). In this context it is interesting to
note that its optical afterglow is particularly faint, R$\sim$23.3 at
t-t$_o$ = 12 hr (Gorosabel et al. 2003), comparable to the dimmest
GRB afterglows found so far.

The fading behavior of the source detected with EPIC, and its
positional coincidence with the
optical transient,  qualify it as the X--ray afterglow of \gr .
Among the afterglows detected so far by \xmm , this one is the brightest and the
one observed with the shortest delay after  the prompt event
(t-t$_{o}$=8 hr).
The EPIC spectrum reported here has the highest statistical
quality among the afterglows detected to date with \xmm .
%, thanks to the brightness
%of the afterglow and the long exposure of this observation.
A power law with a redshifted absorber provides a statistically
acceptable fit to the spectrum.
We found tentative evidence for an emission  line at  1.67 keV.
If this line is due to Fe,
as it has been suggested for similar features observed in other afterglows
(see, e.g., Piro 2002), the
implied redshift of $\sim$2.7-3  (depending on the Fe ionization state)
would be consistent with the value derived from the absorption.
Other interpretations
in terms of lighter elements are of course possible, but somewhat arbitrary
in the lack of an
independent measure of $z$ and/or other statistically significant lines.
% For example, if Fe is instead responsible for the lower significance line
% at 4.37 keV, one could interpret the 1.67 keV line as S XVI K$\alpha$
% for $z$=0.56.
%X--ray data with higher statistical quality are needed to
%Other factors, besides the time elapsed from the prompt event  are
%probably  playing a role for what concerns the presence of a
%detectable thermal component in GRB afterglows.

Low energy emission lines found
in the initial part (t-t$_{o}\sim$11-13 hr) of the GRB 011211 afterglow
(Reeves et al. 2002, but see also Borozdin and Trudolyubov 2003)
provide evidence for thermal emission from a plasma enriched in metals.
Evidence for thermal emission in two other    afterglows
observed with \xmm,
GRB 010220 (t-t$_{o}\sim$15 hr) and GRB 001025A (t-t$_{o}\sim$45 hr),
was reported by Watson et al. (2002a).
On the other hand, the \xmm spectra of the afterglow of
GRB 020322 starting at t-t$_{o}\sim$15 hr (Watson et al. 2002b)
were adequately fit with a power law.
These results suggest that the presence of  thermal components
is not an ubiquitous property of all   GRBs and/or it
might be related only to some short duration phases of the afterglows.

%Our best fit spectrum requires in any case  $z>$1, supporting a non local origin of \gr.
An upper limit to the redshift of $z\leq$3.5 has been
estimated from the absence of the onset of the Lyman forest blanketing
in the optical data (Castro-Tirado et al. 2003b). This allows us
to constrain the $z$ values derived by the X--ray spectral fitting (see Fig. 5),
which is in any case $>$1, supporting a non local
origin of \gr.
The corresponding value of the local absorption  N$_{Hz}$  of the
order of a few times 10$^{22}$ cm$^{-2}$ supports the scenarios involving the occurrence
of GRBs in regions of star formation (e.g. Galama \& Wijers 2001).

Finally we note that
the rapid IBAS localization of \gr leading to the successful detection of its X--ray and
optical afterglows,
as well as the rate of $\sim$1 GRB per month in the IBIS field of view observed so far,
demonstrate that INTEGRAL will efficiently complement
other satellites specifically devoted to GRB studies. We also expect that
particularly interesting results will be obtained for  a few
bursts falling  in the central part of the field of view, which is covered also by
the INTEGRAL monitors operating in the optical (V band) and in the X-ray range
(4.5 - 35 keV).

We thank the \xmm staff for the prompt execution of the TOO observation.
The IBAS development has been funded by the Italian Space Agency.
This research has been partially supported by the  Spanish Program AYA2002-0802
(including FEDER funds) and by Polish grant 2P03C00619p02 from KBN.

\clearpage

\figcaption{Light curve of \gr in the energy range 15 - 300 keV
  obtained with the ISGRI detector of the IBIS instrument.
  The data have been binned in 1 s intervals.
  The two small gaps at t$\sim$10 and 20 s are   artifacts due to
  telemetry saturation.}

\figcaption{Power law index as a function of time for IBIS (triangles)  and SPI (circles).
  Error bars are at 1 $\sigma$.  }

\figcaption{\xmm light curve of the \gr afterglow. All the points are from the
  PN camera, except for the first one obtained with the MOS
  (the remaining MOS points are consistent with the
 PN ones and have not been plotted for clarity).
 The line is a fit with a power law function F$\propto$t$^{-\delta}$ with $\delta$=0.97.  }

\figcaption{ EPIC PN best fit spectra of the afterglow of \gr .}
%  Upper curve (red) refers to the first time interval (Feb 27 18:27-20:32 UT).
%  Lower curve (black) to the second time interval (Feb 27 21:32 UT - Feb 28 06:07 UT). }

\figcaption{ Confidence contours  (68\%, 90\% and 99\% c.l.)
  of redshift and column density of the   redshifted absorber from the fit of the
  whole PN data set. }

\newpage
\epsscale{.7}
\plotone{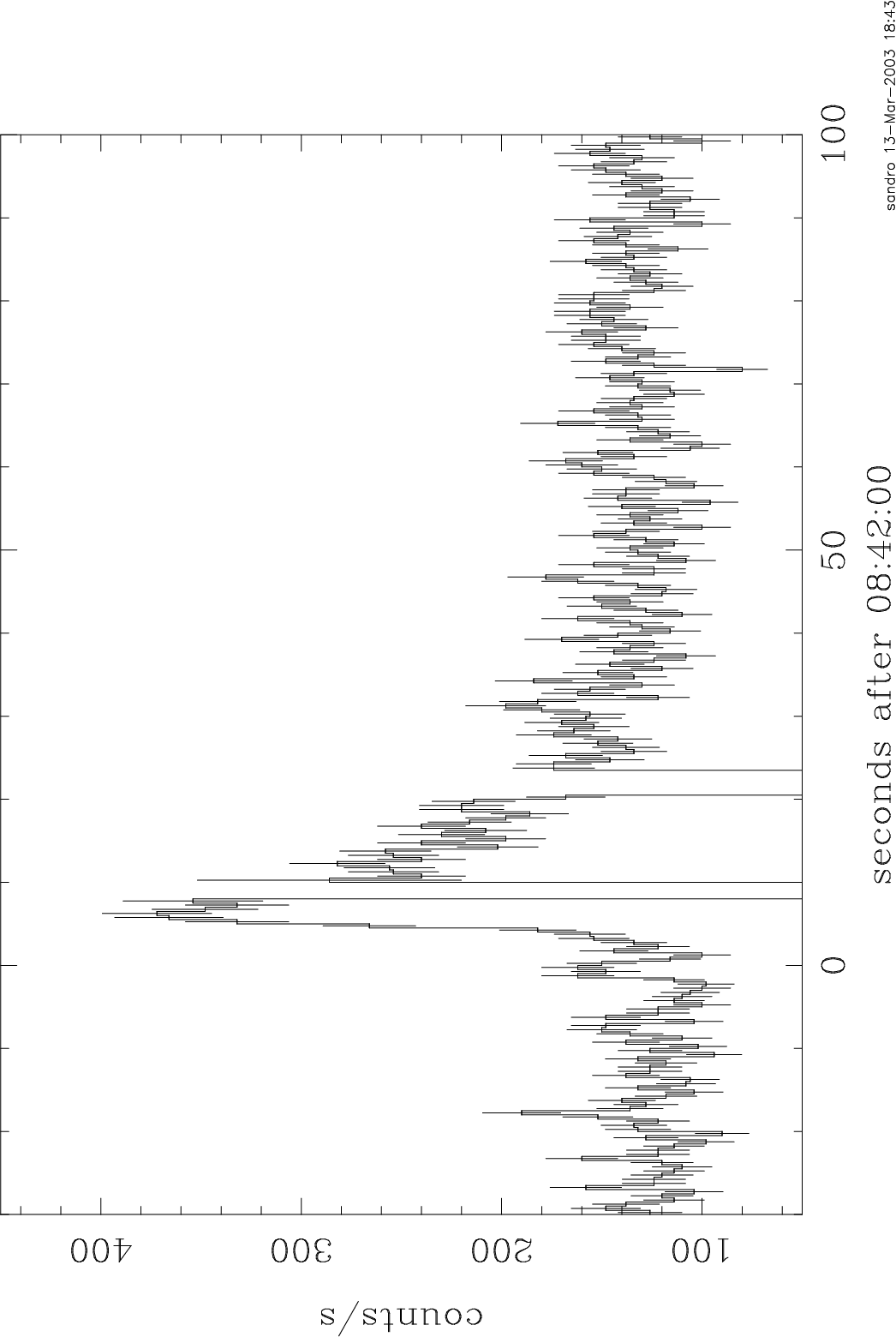}

\newpage
\epsscale{.8}
\plotone{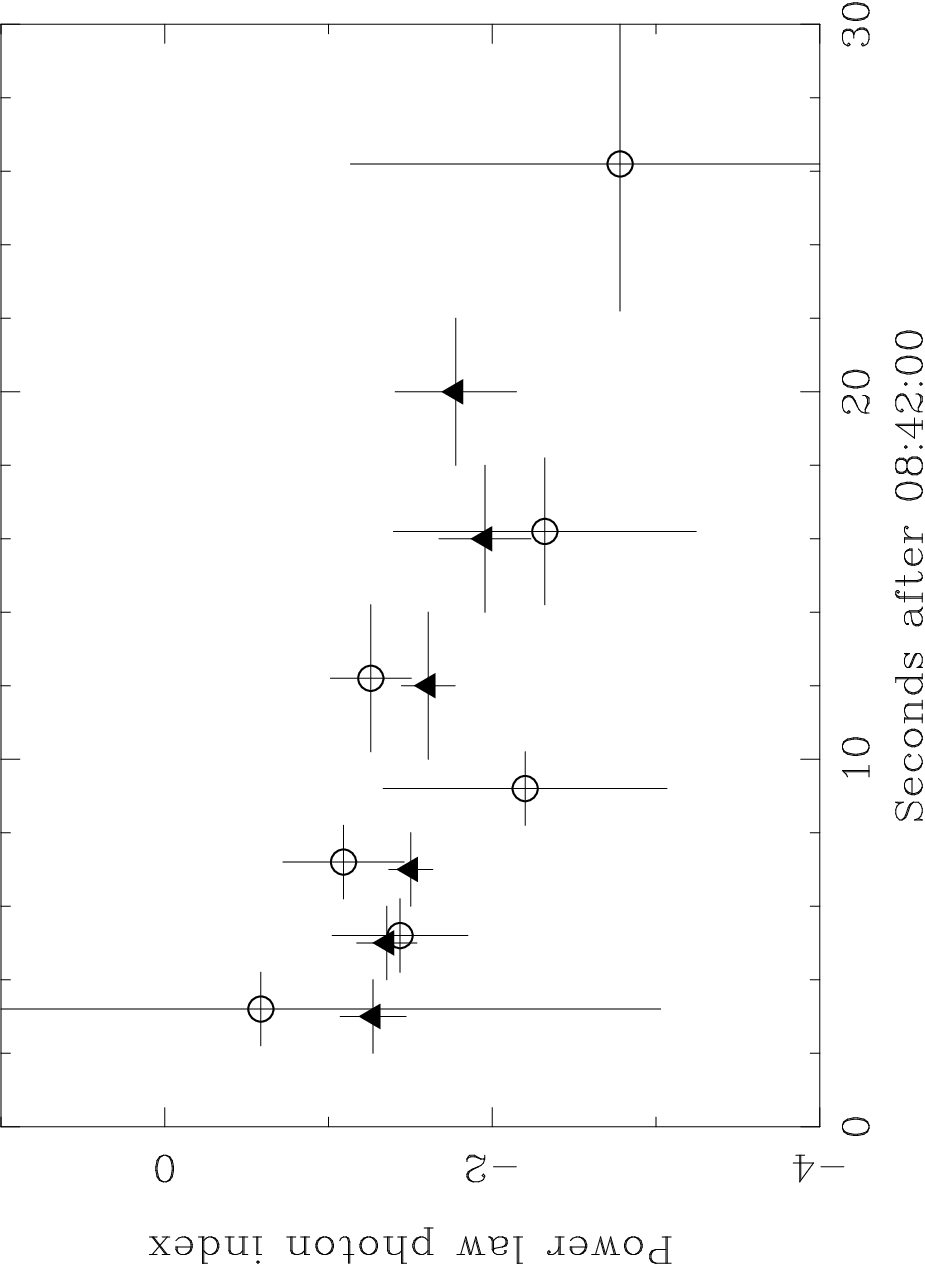}

\newpage
\epsscale{.8}
\plotone{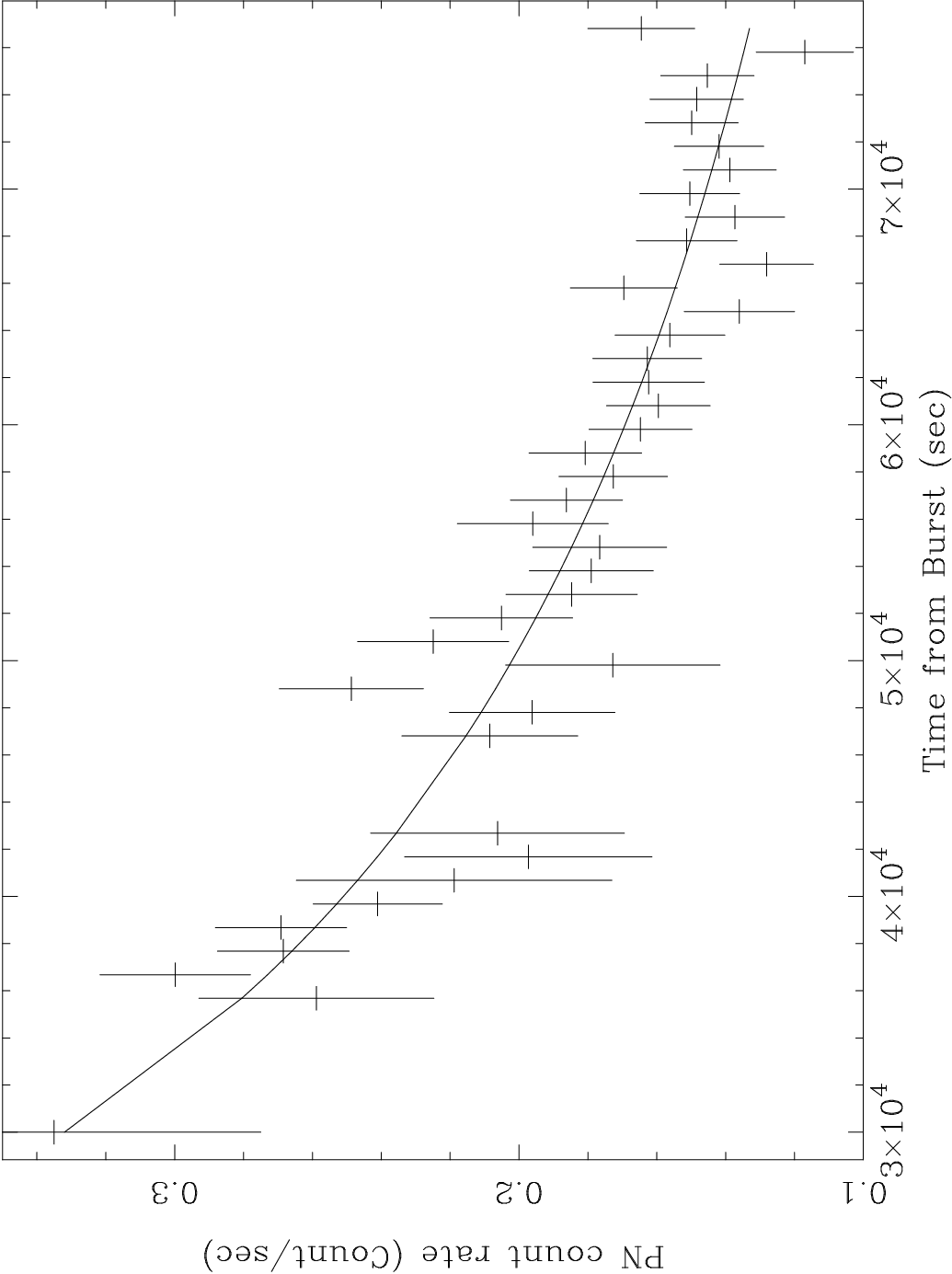}

\epsscale{.8}
\plotone{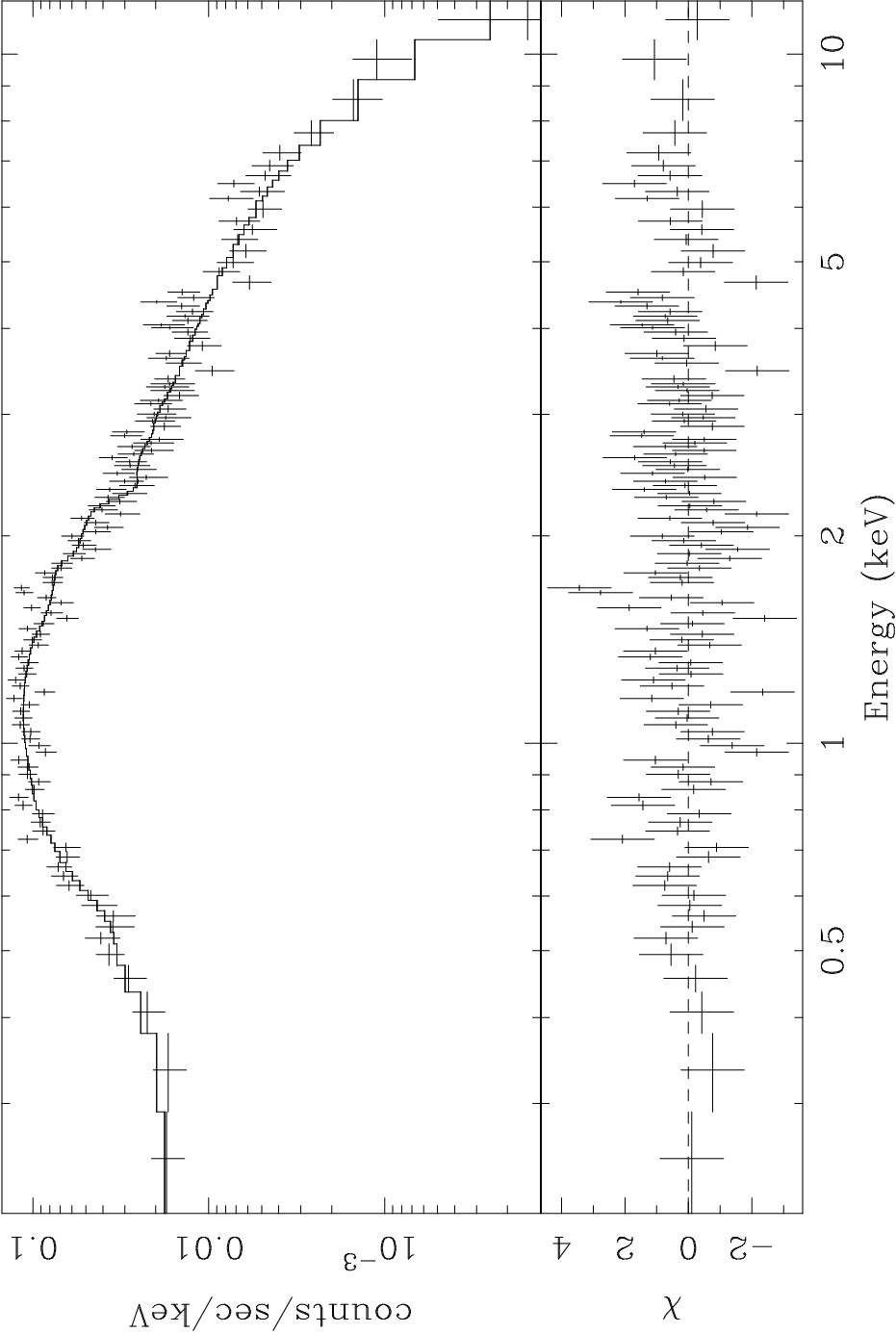}

\newpage
\epsscale{.8}
\plotone{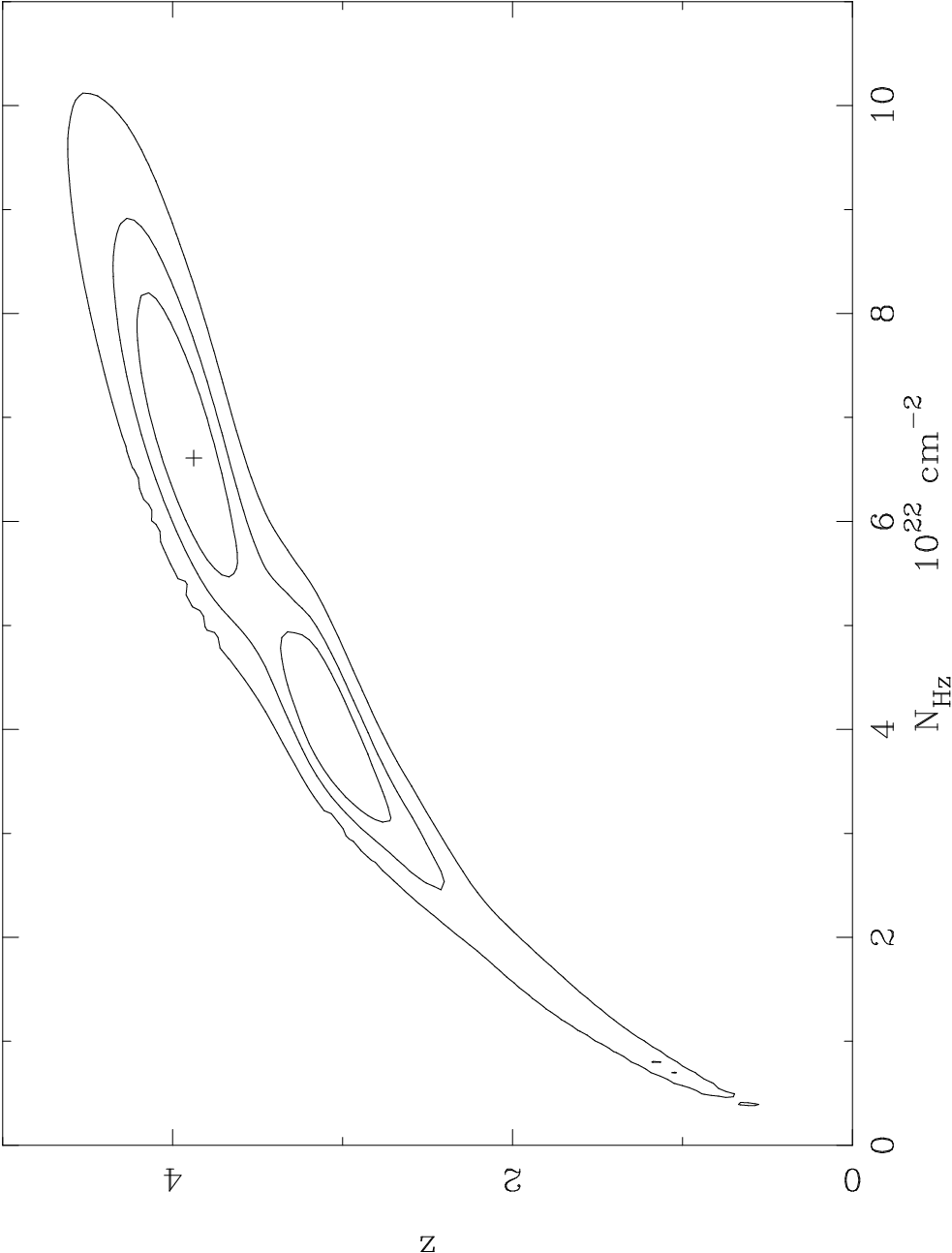}

\end{document}